\documentclass[preprint,1p]{elsarticle}
\journal{Nuclear Physics A}

\usepackage{graphicx,color,amsmath,amssymb}
\usepackage{bm}
\usepackage[colorlinks=true, pdfstartview=FitV, linkcolor=red, citecolor=blue, urlcolor=blue]{hyperref}

\begin{document}

\begin{frontmatter}

\title{ Hydrodynamic approach to p-Pb}

\author[agh,ifj]{Piotr Bo\.zek}
\ead{piotr.bozek@ifj.edu.pl}

\author[ifj,ujk]{Wojciech Broniowski}
\address[agh]{AGH University of Science and Technology, 
Faculty of Physics and Applied Computer Science, PL-30059 Krak\'ow, Poland}
\address[ifj]{
Institute of Nuclear Physics PAN, PL-31342 Krak\'ow, Poland}
\address[ujk]{Institute of Physics, Jan Kochanowski University, PL-25406~Kielce, Poland}
\ead{wojciech.broniowski@ifj.edu.pl}

\begin{abstract}

The formation and  collective expansion of the fireball formed 
in ultrarelativistic p-A and d-A collisions is discussed. Predictions 
of the hydrodynamic model are compared to recent experimental results.
 The presence of strong final state interaction effects in the small dense systems is
 consistent with the observed azimuthal anisotropy of the flow and with the 
mass dependence of the average transverse momentum and of the elliptic flow.
This raises the question of the mechanism explaining such a rapid build up of
the collective flow and the large degree of local equilibration  needed to 
justify this scenario.

\end{abstract}

\end{frontmatter}

\section{Introduction}

Experiments on heavy-ion collisions are performed to study the properties of
  matter at extreme densities. In the interaction region a droplet of hot
and dense matter is formed. If the matter
evolves close to thermal equilibrium, its properties can be deduced 
from a careful comparison to hydrodynamic model calculations.  % \cite{Florkowski:2010zz}.  
Hydrodynamic 
expansion of the  fireball forms the collective flow. The azimuthally asymmetric
collective flow is evidenced in the flow coefficients of final
 particle distributions. % \cite{Heinz:2013th,Gale:2013da,Luzum:2013yya}.
 The study of jet quenching and heavy quark dynamics in the quark gluon plasma 
is based on reference  data that can be obtained in p-A interactions
\cite{Salgado:2011pf}. Assuming that final state interactions are negligible 
in p-A interaction, differences between particle production in A-A 
and p-A that cannot be accounted for by  a simple superposition of
 nucleon-nucleon interactions could 
serve as  probes of the high density quark-gluon plasma.

\begin{figure}
\includegraphics[angle=0,width=0.520 \textwidth]{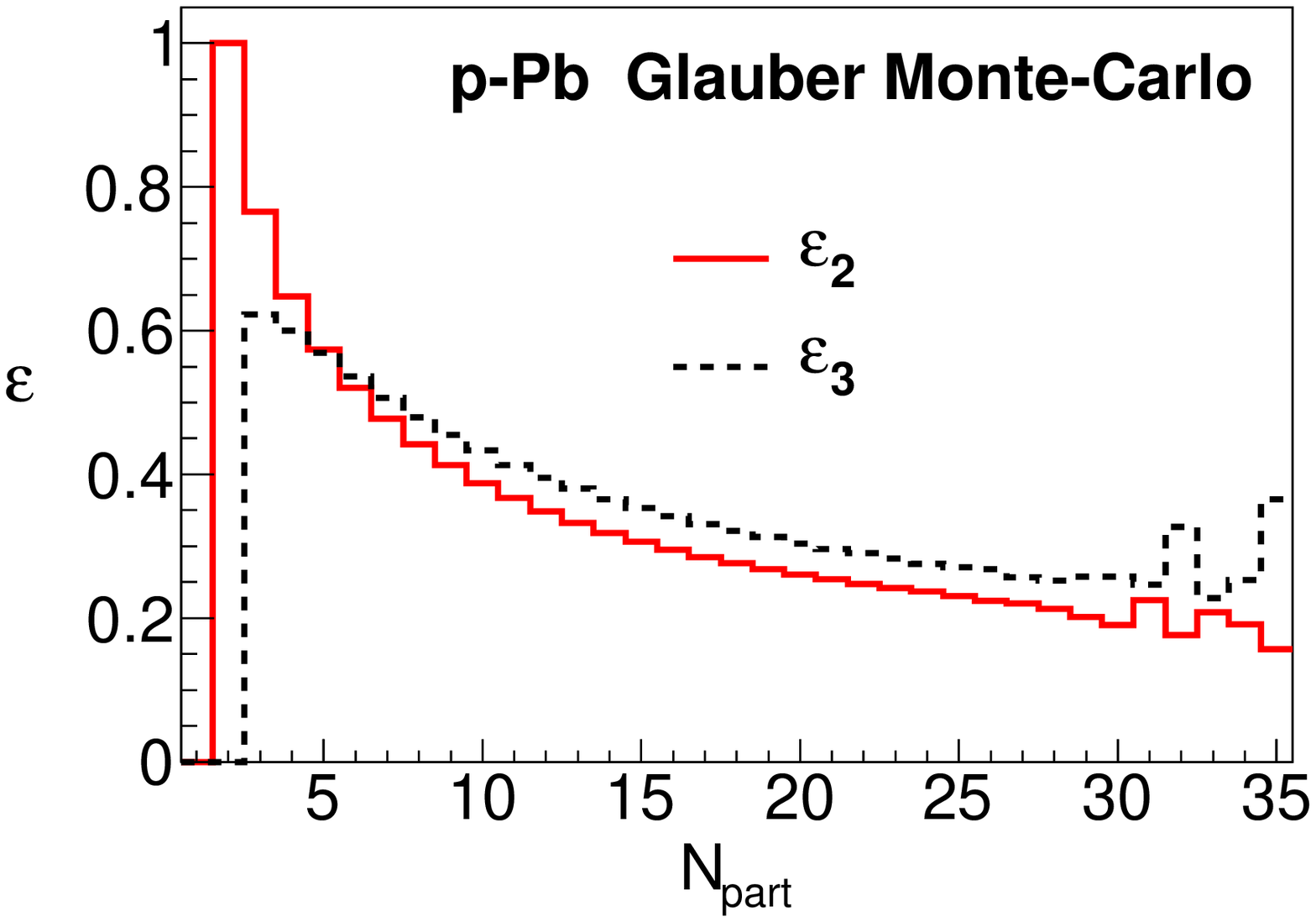}
\includegraphics[angle=0,width=0.480 \textwidth]{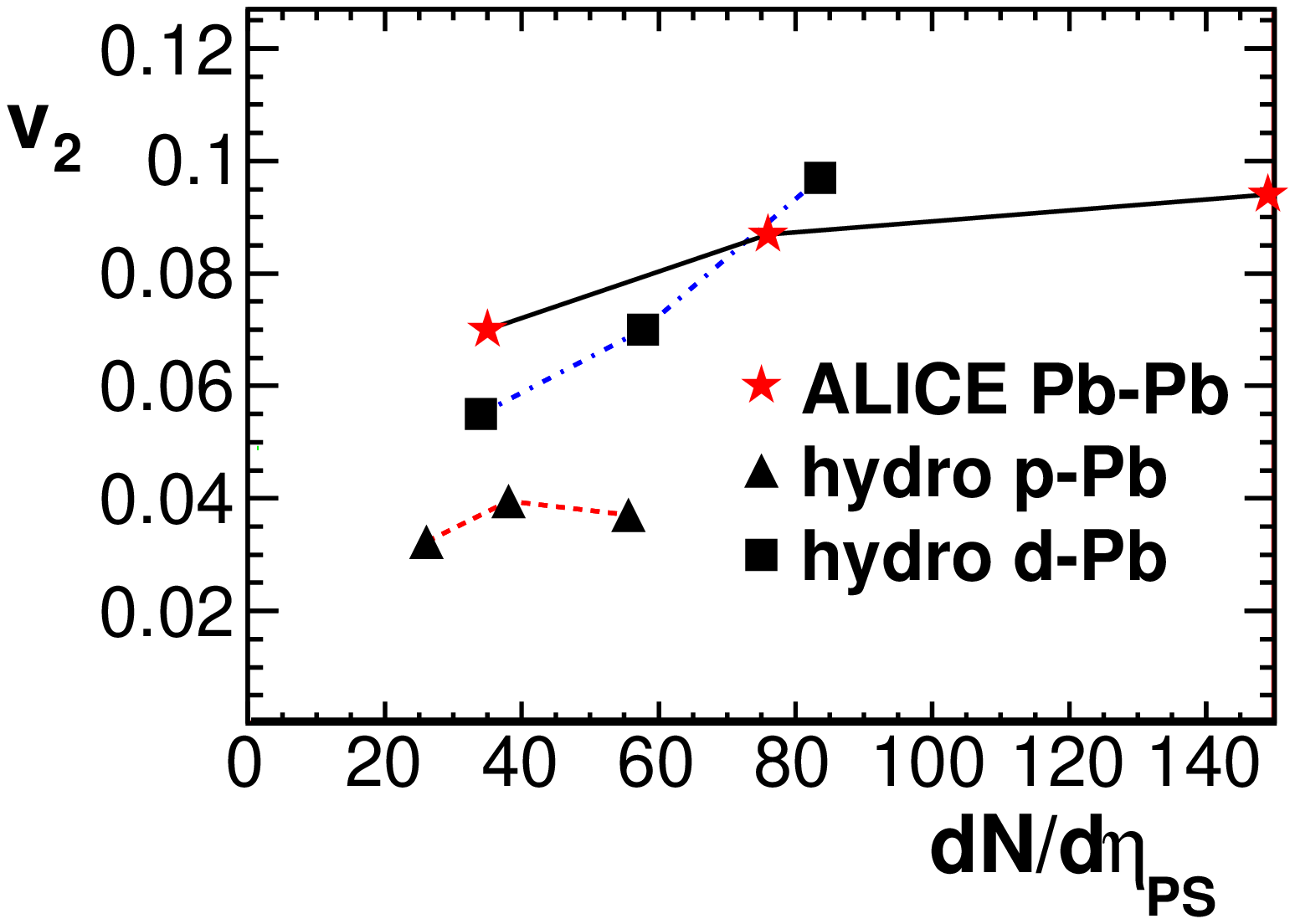} 
\caption{(left panel) Eccentricity and triangularity in p-Pb interactions. 
 (right panel) Predicted 
elliptic flow in p-Pb and d-Pb collisions compared to experimental results
from peripheral Pb-Pb collisions.
\label{fig:eps}} 
\end{figure}

On the other hand, an estimate of the expected particle multiplicity 
in central 
p-Pb interactions at the LHC gave a value  similar as in peripheral
Pb-Pb collisions \cite{Bozek:2011if}. The density of matter created in 
violent p-Pb interactions 
 at the LHC is sufficient for the creation of the quark-gluon 
plasma. The expansion of this
 small fireball leads to noticeable collective flow. 
The size and the shape of the initial source formed in a p-Pb interaction
can be  estimated in the Glauber Monte Carlo model. Fluctuations in the distribution of participant nucleons yield a large eccentricity and triangularity (Fig.~\ref{fig:eps}, left panel). The expansion of the fireball is modeled using  
3+1-dimensional viscous hydrodynamics
 \cite{Schenke:2010rr,Bozek:2011ua}. It has been has been predicted that the
elliptic and triangular flow of particles emitted in central p-Pb collisions 
is large and could be directly measured \cite{Bozek:2011if}.

\begin{figure}
\includegraphics[angle=0,width=0.520 \textwidth]{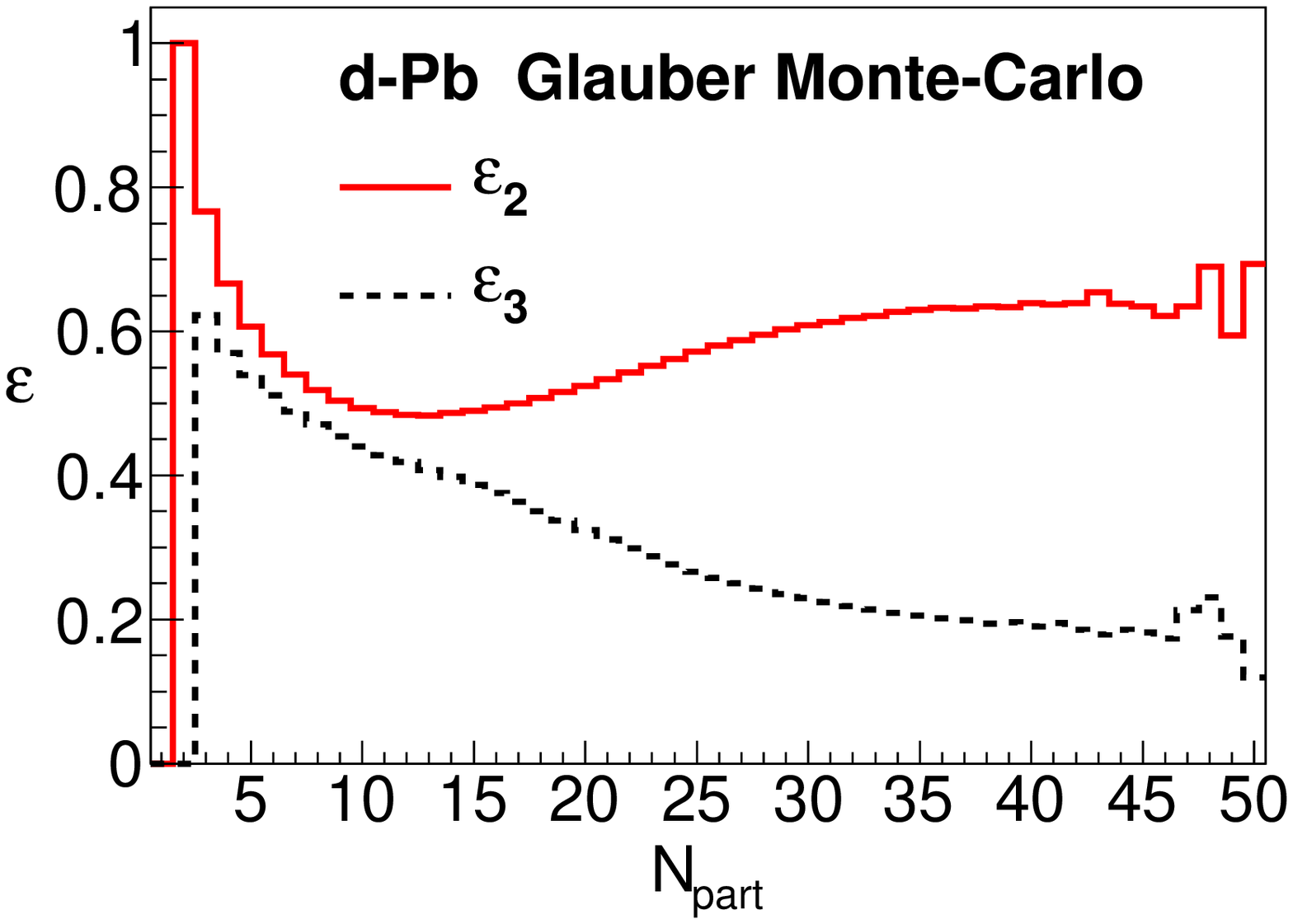}~~~~~~~~~~~~
\includegraphics[angle=0,width=0.335 \textwidth]{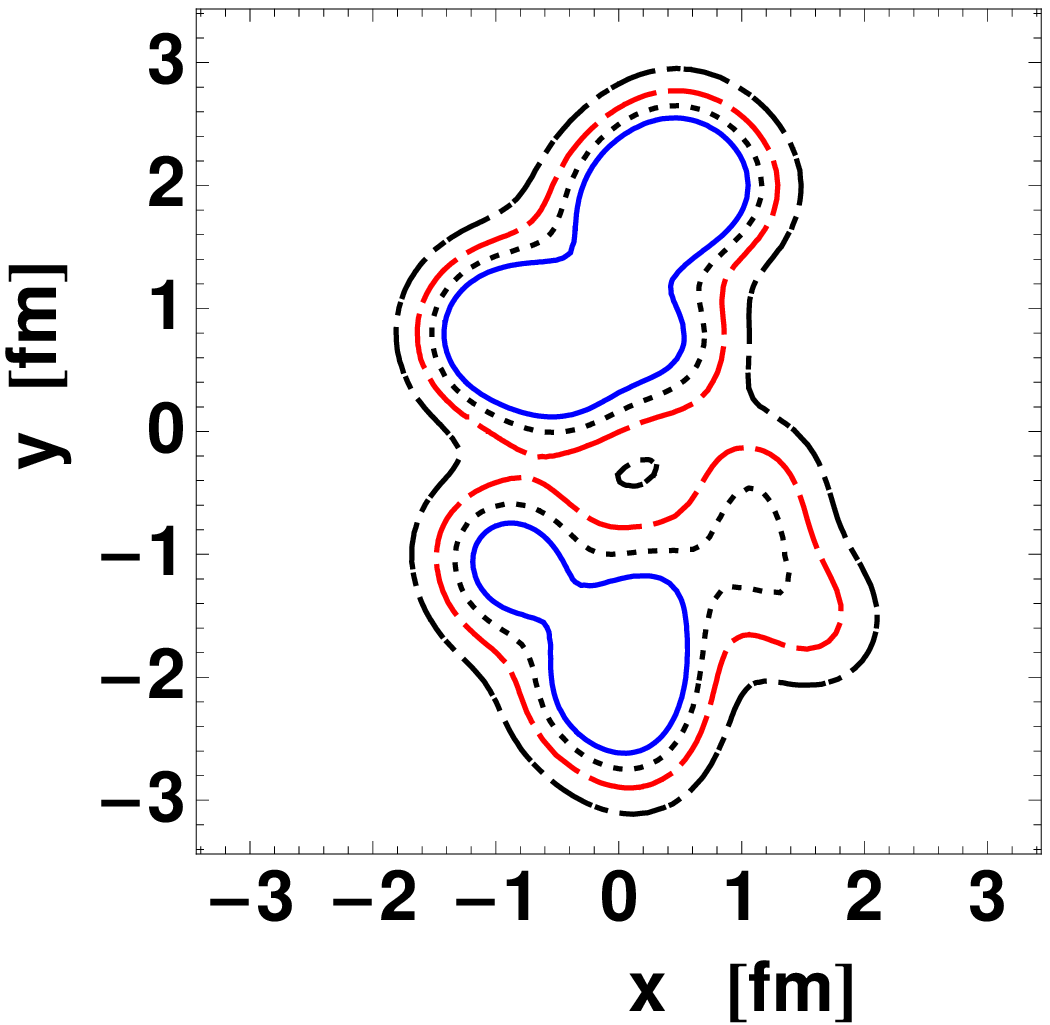} 
\caption{(left panel) Eccentricity and triangularity in d-Pb interactions.
 (right panel) Shape of the d-Pb fireball in the transverse plane.
\label{fig:deps}} 
\end{figure}

An interesting possibility is to study the d-A interactions \cite{Bozek:2011if}.
This is a small dense system, similar as for  the p-A case, but the shape
 of the interaction region is very much deformed (Fig. \ref{fig:deps}, right panel).
The most violent collisions, corresponding to a large number of participants, 
happen when the deuteron hits the large nucleus side-wise. The eccentricity of the fireball can be as large as $0.5$ (Fig. \ref{fig:deps}, left panel). Moreover, large value of the  initial
 eccentricity is defined by the geometry of the projectile and is not much influenced by the 
fluctuations. Experimentally, such configurations can be triggered on by choosing events with 
high multiplicity (transverse energy).
The prediction  is that in the most violent d-A collisions the elliptic flow 
is very large \cite{Bozek:2011if}, and could be larger than in p-A 
or peripheral A-A collisions (Fig. \ref{fig:eps}, right panel).

\section{Hydrodynamic flow in p-Pb}

The analysis of two-particle correlations in the relative azimuthal
 angle and relative pseudorapidity for p-Pb collisions at the LHC revealed
the presence of  long-range ridge like structures for particle
pairs emitted in the same ($\Delta \phi \simeq 0$) and away
 side ($\Delta \phi \simeq \pi$) directions 
\cite{CMS:2012qk,Abelev:2012ola,Aad:2012gla}.
The ridge-like structures are very similar as observed in A-A and in high multiplicity  p-p collisions
\cite{Wenger:2008ts,Agakishiev:2011pe,Chatrchyan:2011eka,Khachatryan:2010gv}.
\begin{figure}
\begin{center}
\includegraphics[angle=0,width=0.700 \textwidth]{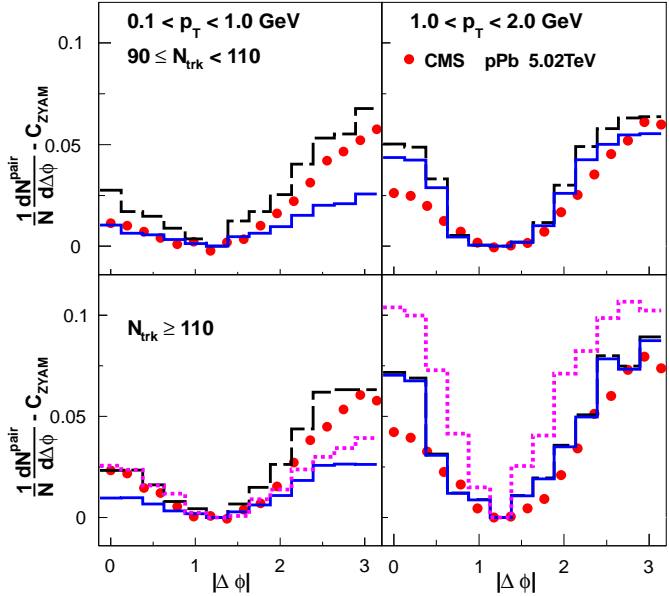}
\end{center}
\caption{The   zero yield at minimum subtracted correlation function in relative azimuthal angle  in  p-Pb collisions. 
The CMS measurement~\cite{CMS:2012qk} is shown as dots.  The results 
of our hydrodynamic model with the normalized correlation functions are shown with the solid lines. 
The dashed lines show the results of the hydrodynamic model with subtraction of the model ZYAM 
values and no rescaling. The dotted lines represent the results obtained with a smaller initial time of $0.2$~fm/c.
\label{fig:corrpanel}} 
\end{figure}  
In heavy-ion collisions, but also in p-p interactions, such structures can be interpreted as due to 
collective flow  harmonics $v_{2}$, $v_{3}$,  and to momentum conservation effects \cite{Takahashi:2009na,Luzum:2010sp,Bozek:2010pb}. The two-hadron correlation function projected on the relative angle can be decomposed in successive harmonics
\begin{equation}
C(\Delta \phi)\propto 1 + C_1 \cos(\Delta \phi)+ 2 v_2^2 \cos(2 \Delta \phi)+ 2 v_3^2 \cos(3\Delta \phi)+ \dots \ \ .
\end{equation}
For p-Pb collisions, in the hydrodynamic scenario the coefficients $v_{2}$ and $v_{3}$ are the  elliptic and triangular flow coefficients, while $C_1<0$ is due mainly to transverse momentum conservation
 effects at finite multiplicity. In Fig. \ref{fig:corrpanel} the two-hadron
 correlation function measured by the CMS Collaboration \cite{CMS:2012qk}
is compared to hydrodynamic calculations \cite{Bozek:2012gr}.
The  collective expansion scenario can explain semi-quantitatively the observed structures.
The same-side ridge in the correlation function normalized by the number of trigger particles 
increases with multiplicity. This dependence is natural for collective flow correlations 
which correlate all the particles in the event. 

We note, that 
 angular correlations between  emitted 
hadrons can arise from interference diagrams  between  gluons in the initial
 state  
\cite{Dusling:2012iga,Dusling:2012cg,Dusling:2013oia,Dusling:2012wy}.
Such effects are expected in the color glass condensate regime
 and explain the observed small angle 
enhancement of the two-hadron correlation function in p-p and p-Pb collisions, as well as its $p_\perp$ 
dependence.

\begin{figure}
\begin{center}
\includegraphics[angle=0,width=0.600 \textwidth]{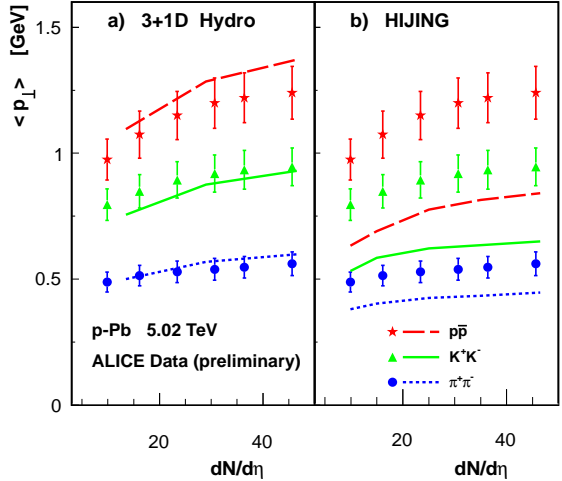}
\end{center}
\caption{Average transverse momentum of pions, kaons, and protons in p-Pb collisions measured by the 
ALICE Collaboration \cite{Abelev:2013haa} compared to results of the  HIJING model (right panel) and 
 of the hydrodynamic model (left panel). 
\label{fig:pt}} 
\end{figure}

In the hydrodynamic picture the transverse momentum of an emitted particle has a component from the 
collective flow of the fluid. The stronger is the generated transverse flow, the larger is 
$\langle p_\perp \rangle$. Another effect of the collective transverse flow is
 the mass hierarchy of the
transverse momenta, i.e., $\langle p_\perp \rangle$ increases with the particle mass. Both effects 
are observed in p-p,  p-Pb and Pb-Pb collisions 
\cite{Abelev:2013bla,Abelev:2013haa,Chatrchyan:2013eya}. 
The increase of the average transverse momentum with the event multiplicity and its 
mass hierarchy in p-p interactions can be explained by the color reconnection mechanism 
\cite{Ortiz:2013yxa}. However, in
  p-Pb collisions,  $\langle p_\perp \rangle$ 
 is stronger than expected from a  superposition model \cite{Bzdak:2013lva}.
An example of a superposition model is given by the HIJING model calculation shown in Fig. 
\ref{fig:pt}, where the calculated transverse momentum is smaller than observed
 and its dependence on the particle mass is too weak. The  measured transverse momenta can be explained as due to 
the collective transverse flow
 generated in the hydrodynamic expansion of the fireball 
formed in p-Pb collisions (Fig. \ref{fig:pt}, left panel) \cite{Bozek:2013ska,Shuryak:2013ke,Bzdak:2013lva,Qin:2013bha,Werner:2013ipa}. 
An alternative scenario based on
 geometrical scaling leads to a mass hierarchy in  $\langle p_\perp \rangle$ as well
\cite{McLerran:2013oju}.

\begin{figure}
\includegraphics[angle=0,width=0.5 \textwidth]{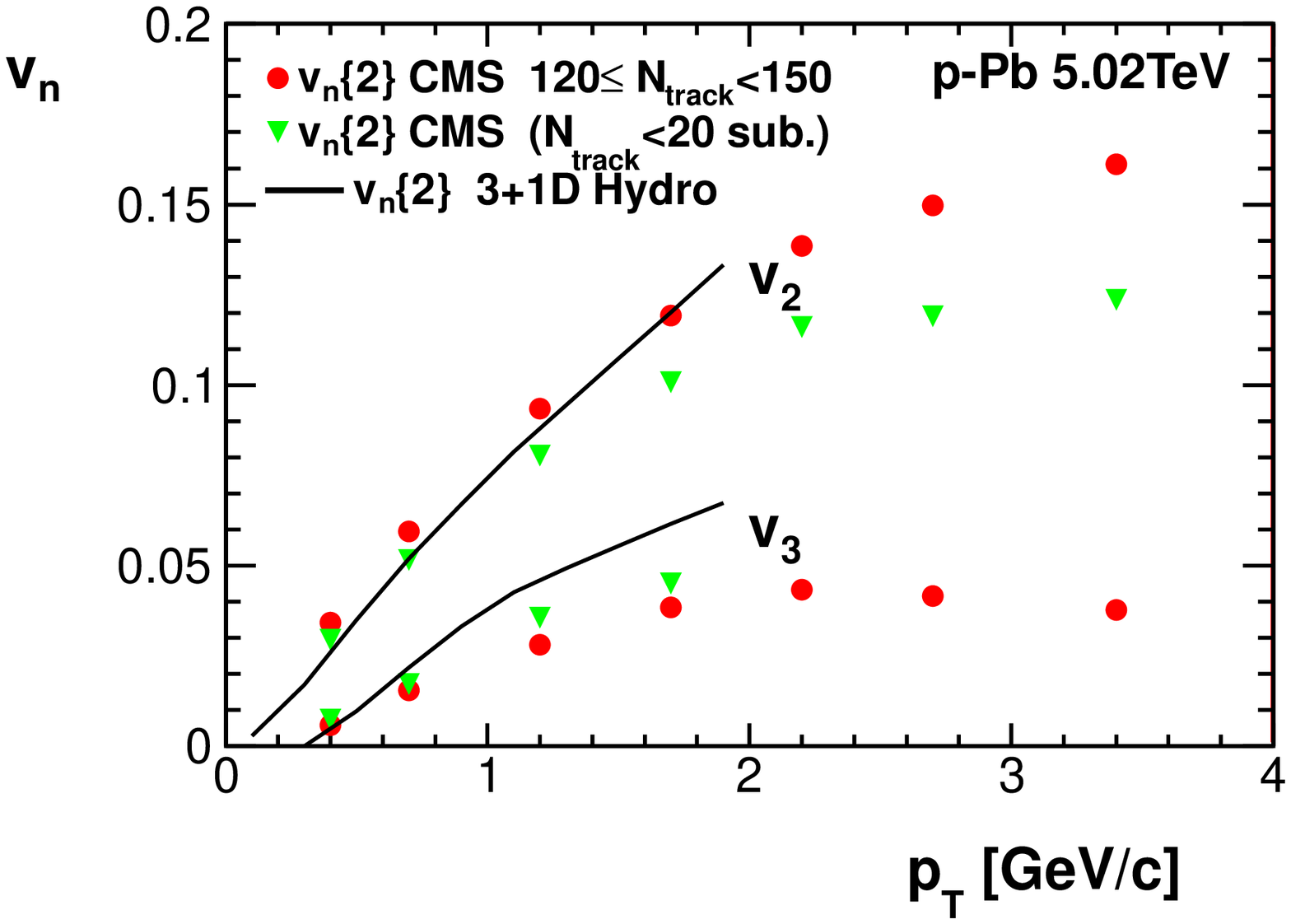}
\includegraphics[angle=0,width=0.5 \textwidth]{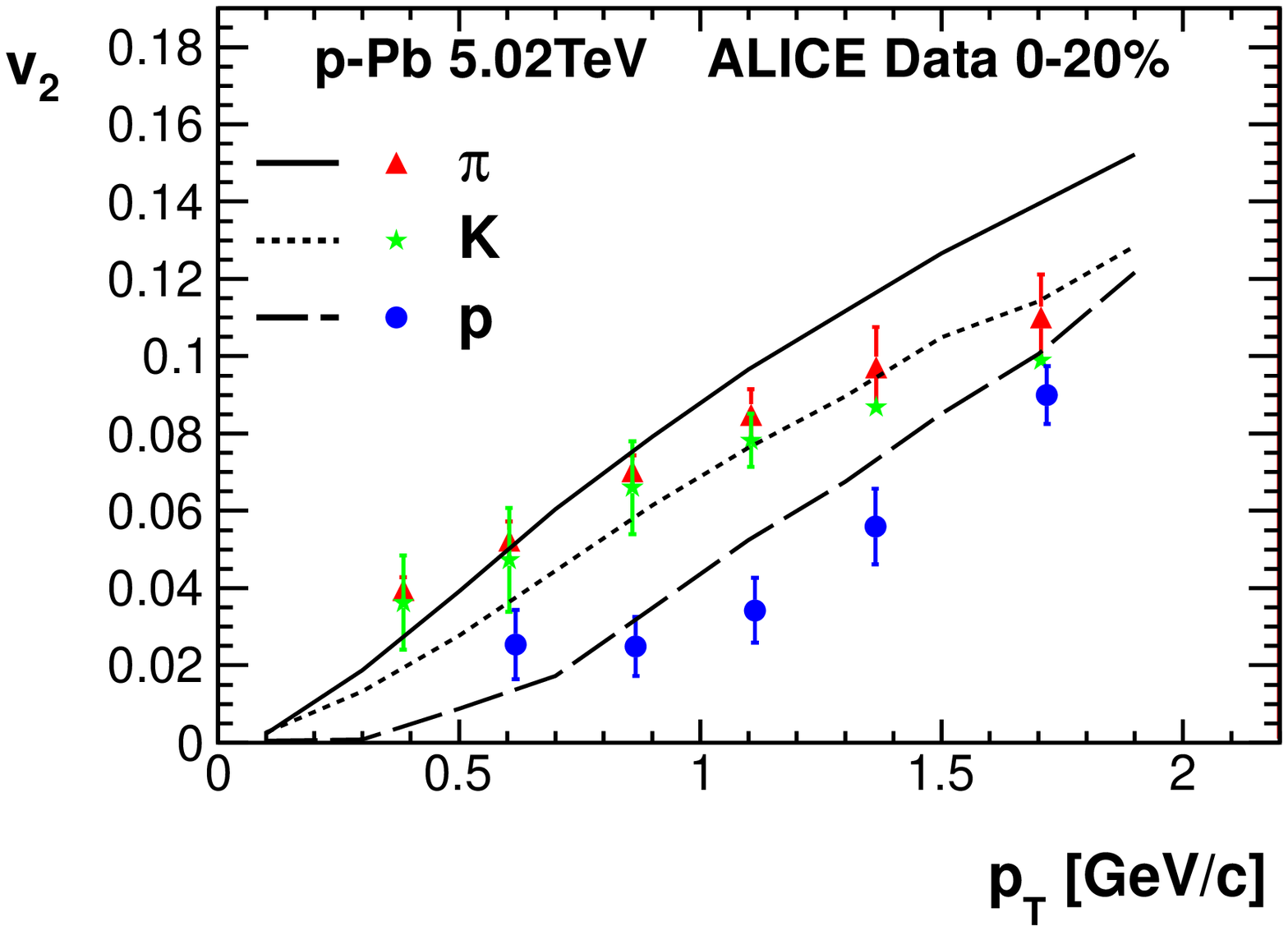} 
\caption{(left panel) Calculated 
elliptic and triangular flow coefficient of charged particles in p-Pb collisions compared 
to the CMS data \cite{Chatrchyan:2013nka}.
 (right panel) Calculated elliptic flow coefficient of identified particles compared to the ALICE 
data \cite{ABELEV:2013wsa}.
\label{fig:v23}} 
\end{figure}  

The flow coefficients $v_2$ and $v_3$ can be measured in p-Pb collisions at
 the LHC \cite{Aad:2013fja,Chatrchyan:2013nka,ABELEV:2013wsa}. 
The procedure is more involved than in central A-A collisions 
because of significant nonflow correlations, from jets, resonance decays,
 or momentum conservation. Some of the nonflow correlations can be reduced 
using a rapidity gap between the particles or using the 
fourth order cumulant  \cite{Borghini:2000sa,Xu:2012ue}. Another possibility is to 
 subtract the per-trigger two-particle correlation for peripheral events 
from the one obtained for central events. Significant elliptic
 and triangular flow coefficients are measured in central p-Pb collisions. 
The outcomes of different methods of reducing nonflow effects vary somewhat, 
but  correlations consistent 
 with collective flow are observed within the uncertainty of the procedure.

In Fig. \ref{fig:v23} the results of a hydrodynamic calculation
 are compared to experimental data on flow coefficients
for charged and identified particles \cite{Bozek:2013ska}.
The $p_\perp$ dependent flow coefficients for charged 
particles for central p-Pb collisions are well reproduced by our hydrodynamic model.
Qualitatively similar results are obtained for other calculations
using different assumptions on the initial density and the fluid properties
\cite{Bozek:2013uha,Bzdak:2013lva,Qin:2013bha}. The magnitude of the predicted 
flow depends on the details of the model. The centrality dependence of the 
$v_2$ and $v_3$ coefficients cannot be fully reproduced in the hydrodynamic 
calculations. When varying the initial size of the fireball 
from peripheral to central p-Pb collisions, the collective expansion 
 sets in. However, it is difficult to model accurately
this transition using viscous hydrodynamics, which assumes almost complete equilibration. The harmonic flow coefficients have a mass hierarchy. For 
soft momenta, the $v_2$ coefficient for pions is larger than for protons
(Fig. \ref{fig:v23}, right panel).
The fluid dynamic calculations reproduce this effect
\cite{Bozek:2013ska,Werner:2013ipa}.

The d-Au at RHIC energies have been analyzed by the PHENIX Collaboration
 \cite{Adare:2013piz}. The peripheral from central events subtraction 
procedure has been applied to the two-hadron correlation functions.
The nonflow effects are relatively more important than for p-Pb collisions at the LHC, since the average multiplicities are lower. The extracted elliptic 
flow coefficient is large, in line with what has been predicted in the hydrodynamic model \cite{Bozek:2011if}. The intrinsic 
deformation of the deuteron projectile 
leads to a very large eccentricity in the initial state, but the triangularity 
is determined by fluctuations. The final $v_2$ coefficient is large and
$v_3$ is small \cite{Bozek:2011if,Bzdak:2013lva,Qin:2013bha}. Further 
insight on
the collectivity in small systems could be gained from $^3$He-Au collisions,
involving a projectile with triangular deformation \cite{Sickles:2013mua}.
Interestingly, deformations due to alpha clustering in small nuclei could also be  studied in relativistic nuclear collisions~\cite{Broniowski:2013dia}.

The initial size of the fireball in p-Pb collisions is expected 
to be smaller than in
 peripheral Pb-Pb collisions. The final size of 
the system can be measured using the
 Hanbury Brown-Twiss   correlation 
radii. % \cite{Wiedemann:1999qn,Lisa:2005dd}. 
If the system undergoes as substantial collective expansion 
phase, its size at freeze-out would be comparable to the freeze-out radius 
of the fireball in peripheral Pb-Pb collisions, 
otherwise one expects to find somewhat smaller 
femtoscopic radii \cite{Bozek:2013df,Bzdak:2013lva}. 
\begin{figure}
\begin{center}
\includegraphics[angle=0,width=0.700 \textwidth]{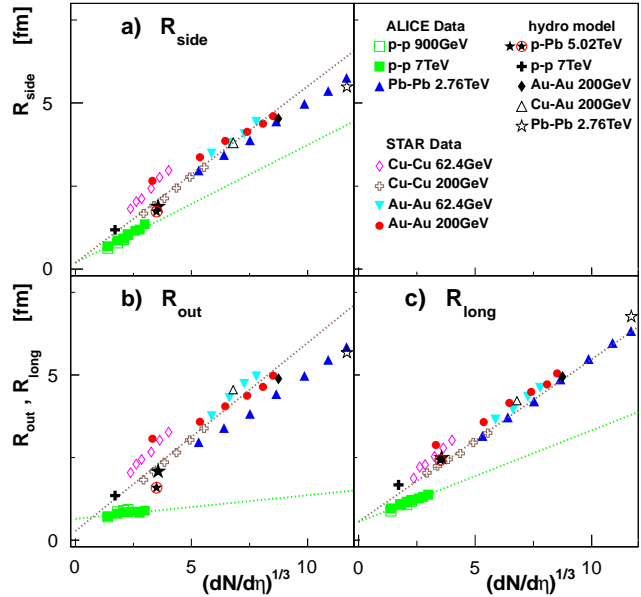}
\end{center}
\caption{Pion interferometry radii $R_{\rm side}$ (a), $R_{\rm out}$ (b), and $R_{\rm long}$ (c)
 for different collisions systems
as functions of the  multiplicity~\cite{Kisiel:2011jt}. % ~\cite{Kisiel:2011jg,Kisiel:2011jt}. 
%For RHIC energies we quote the STAR Collaboration results~\cite{Abelev:2009tp,Adams:2004yc} 
% for the LHC energies  the ALICE Collaboration data
%\cite{Aamodt:2011kd,Aamodt:2011mr}. 
The  hydrodynamic calculations are given for high multiplicity  p-p, Cu-Au, Pb-Pb, and p-Pb 
collisions. % ~\cite{Bozek:2009dt,Bozek:2012hy,Bozek:2011ua,Bozek:2013df}. 
\label{fig:hbtsys}} 
\end{figure}  
Explicit calculations in 
 the hydrodynamic model for p-Pb collisions show that the predicted 
femtoscopy radii are similar 
as in the A-A collisions with similar multiplicity (Fig. \ref{fig:hbtsys}).

\section{Collectivity in small systems}

The hydrodynamic model describes in a satisfactory way a number of experimental observations in 
p-Pb collisions. This raises the question, whether 
for such a small and short living system the 
assumption of approximate local equilibration is justified. Within  hydrodynamics 
small deviations from equilibrium can be quantified as viscosity corrections. 
While the local energy density in a high multiplicity  p-Pb event  is above
 the phase transition density, the small 
size of the fireball leads to large  velocity gradients. 
The mean free path is not very small compared to the system size and the viscous corrections
 in the dynamics are large \cite{Danielewicz:1984ww,Liao:2009gb,Bzdak:2013lva}.
These arguments shows that the agreement of the hydrodynamic calculations with the data
 for p-Pb can be at best semi-quantitative.

At the early phase of the collision, corrections to 
the energy-momentum tensor of a perfect fluid amount to a strong reduction of the
 longitudinal pressure and an increase of the transverse pressure
\cite{Florkowski:2013sk,Heller:2011ju}. The longitudinal pressure can be even negative.
This  seems to invalidate any hydrodynamic approach. However, universal flow arguments 
 and numerical simulations 
\cite{Vredevoogd:2008id,Broniowski:2008qk,Bozek:2010aj,Ryblewski:2012rr} 
show that the transverse 
collective flow generated in the early, far 
from equilibrium evolution is almost the same as the flow generated
 in a perfect fluid expansion starting from the same initial time.
This means that hydrodynamics acts as an effective theory for the transverse 
dynamics of the  system even if the system is far from equilibrium.

From the arguments given  above one should not conclude that {\it any} dynamical model
with  gradients of the energy-momentum tensor
would give the observed flow. In order for the universal flow argument to apply, the system 
must eventually undergo a {\it thermalization} stage
 \cite{Vredevoogd:2008id,Broniowski:2008qk}. In fact, only after an approximate 
local thermalization the local flow velocity and energy density are defined.

It is possible that in the short lived p-Pb system most of the evolution occurs far
 from equilibrium, with  the thermalization happening only just before hadronization. 
The final  transverse, elliptic and triangular flow in that scenario cannot
 be 
distinguished from  the quantities obtained in a pure hydrodynamic expansion. 
The dynamics of small systems offers a nice laboratory to test the isotropization mechanism with probes sensitive to the pressure anisotropy.
 Observables involving explicitly the longitudinal pressure have been discussed in that respect, namely, the dilepton production \cite{Mauricio:2007vz}, photon emission \cite{Schenke:2006yp,Bhattacharya:2008mv}, directed flow \cite{Bozek:2010aj}, and bottomonium suppression
\cite{Strickland:2011aa}.

\section*{Acknowledgments}
This work is partly supported by the National Science Centre, 
Poland, grant DEC-2012/05/B/ST2/02528, and PL-Grid infrastructure.

\bibliography{../../../hydr}

\begin{thebibliography}{10}

\bibitem{Salgado:2011pf}
C.A. Salgado,
\newblock J.Phys.G G38 (2011) 124036.

\bibitem{Bozek:2011if}
P. Bo\.zek,
\newblock Phys. Rev. C85 (2012) 014911.

\bibitem{Schenke:2010rr}
B. Schenke, S. Jeon and C. Gale,
\newblock Phys. Rev. Lett. 106 (2011) 042301.

\bibitem{Bozek:2011ua}
P. Bo\.zek,
\newblock Phys. Rev. C85 (2012) 034901.

\bibitem{CMS:2012qk}
CMS Collaboration, S. Chatrchyan et~al.,
\newblock Phys. Lett. B718 (2013) 795.

\bibitem{Abelev:2012ola}
ALICE Collaboration, B. Abelev et~al.,
\newblock Phys. Lett. B719 (2013) 29.

\bibitem{Aad:2012gla}
ATLAS Collaboration, G. Aad et~al.,
\newblock Phys. Rev. Lett. 110 (2013) 182302.

\bibitem{Wenger:2008ts}
PHOBOS Collaboration, B. Alver et~al.,
\newblock J.Phys. G35 (2008) 104080.

\bibitem{Agakishiev:2011pe}
STAR Collaboration, G. Agakishiev et~al.,
\newblock Phys. Rev. C86 (2012) 064902.

\bibitem{Chatrchyan:2011eka}
CMS Collaboration, S. Chatrchyan et~al.,
\newblock JHEP 1107 (2011) 076.

\bibitem{Khachatryan:2010gv}
CMS, V. Khachatryan et~al.,
\newblock JHEP 09 (2010) 091.

\bibitem{Takahashi:2009na}
J. Takahashi et~al.,
\newblock Phys. Rev. Lett. 103 (2009) 242301.

\bibitem{Luzum:2010sp}
M. Luzum,
\newblock Phys. Lett. B696 (2011) 499.

\bibitem{Bozek:2010pb}
P. Bo\.zek,
\newblock Eur.Phys.J. C71 (2011) 1530.

\bibitem{Bozek:2012gr}
P. Bo\.zek and W. Broniowski,
\newblock Phys. Lett. B718 (2013) 1557.

\bibitem{Dusling:2012iga}
K. Dusling and R. Venugopalan,
\newblock Phys. Rev. Lett. 108 (2012) 262001.

\bibitem{Dusling:2012cg}
K. Dusling and R. Venugopalan,
\newblock Phys. Rev. D 87 (2013) 051502.

\bibitem{Dusling:2013oia}
K. Dusling and R. Venugopalan,
\newblock Phys. Rev. D87 (2013) 094034.

\bibitem{Dusling:2012wy}
K. Dusling and R. Venugopalan,
\newblock Phys. Rev. D 87 (2013) 054014.

\bibitem{Abelev:2013haa}
ALICE Collaboration, B.B. Abelev et~al.,
\newblock (2013), 1307.6796 [nucl-ex].

\bibitem{Abelev:2013bla}
ALICE Collaboration, B.B. Abelev et~al.,
\newblock (2013), 1307.1094 [nucl-ex].

\bibitem{Chatrchyan:2013eya}
CMS Collaboration, S. Chatrchyan et~al.,
\newblock (2013), 1307.3442 [hep-ex].

\bibitem{Ortiz:2013yxa}
A. Ortiz et~al.,
\newblock Phys. Rev. Lett. 111 (2013) 042001.

\bibitem{Bzdak:2013lva}
A. Bzdak and V. Skokov,
\newblock Phys. Lett. B726 (2013) 408.

\bibitem{Bozek:2013ska}
P. Bo\.zek, W. Broniowski and G. Torrieri,
\newblock Phys. Rev. Lett. 111 (2013) 172303.

\bibitem{Shuryak:2013ke}
E. Shuryak and I. Zahed,
\newblock Phys.Rev. C88 (2013) 044915.

\bibitem{Qin:2013bha}
G.Y. Qin and B. M{\"u}ller,
\newblock (2013), 1306.3439 [nucl-th].

\bibitem{Werner:2013ipa}
K. Werner et~al.,
\newblock (2013), 1307.4379 [nucl-th].

\bibitem{McLerran:2013oju}
L. McLerran, M. Praszalowicz and B. Schenke,
\newblock Nucl.Phys. A916 (2013) 210.

\bibitem{Chatrchyan:2013nka}
CMS Collaboration, S. Chatrchyan et~al.,
\newblock Phys. Lett. B724 (2013) 213.

\bibitem{ABELEV:2013wsa}
ALICE Collaboration, B.B. Abelev et~al.,
\newblock Phys. Lett. B726 (2013) 164.

\bibitem{Aad:2013fja}
ATLAS Collaboration, G. Aad et~al.,
\newblock Phys. Lett. B725 (2013) 60.

\bibitem{Borghini:2000sa}
N. Borghini, P.M. Dinh and J.Y. Ollitrault,
\newblock Phys. Rev. C63 (2001) 054906.

\bibitem{Xu:2012ue}
L. Xu et~al.,
\newblock Phys.Rev. C86 (2012) 024910.

\bibitem{Bozek:2013uha}
P. Bo\.zek and W. Broniowski,
\newblock Phys. Rev. C88 (2013) 014903.

\bibitem{Adare:2013piz}
PHENIX Collaboration, A. Adare et~al.,
\newblock Phys. Rev. Lett. 111 (2013) 212301.

\bibitem{Sickles:2013mua}
PHENIX Collaboration, A.M. Sickles,
\newblock (2013), 1310.4388 [nucl-ex].

\bibitem{Broniowski:2013dia}
 W.~Broniowski and E.~R.~Arriola,
\newblock (2013), 1312.0289 [nucl-th].

\bibitem{Bozek:2013df}
P. Bo\.zek and W. Broniowski,
\newblock Phys. Lett. B720 (2013) 250.

\bibitem{Kisiel:2011jt}
ALICE Collaboration, A. Kisiel,
\newblock PoS WPCF2011 (2011) 003.

\bibitem{Danielewicz:1984ww}
P. Danielewicz and M. Gyulassy,
\newblock Phys. Rev. D31 (1985) 53.

\bibitem{Liao:2009gb}
J. Liao and V. Koch,
\newblock Phys. Rev. C81 (2010) 014902.

\bibitem{Florkowski:2013sk}
W. Florkowski et~al.,
\newblock (2013), 1301.7539 [nucl-th].

\bibitem{Heller:2011ju}
M.P. Heller, R.A. Janik and P. Witaszczyk,
\newblock Phys. Rev. Lett. 108 (2012) 201602.

\bibitem{Vredevoogd:2008id}
J. Vredevoogd and S. Pratt,
\newblock Phys. Rev. C79 (2009) 044915.

\bibitem{Broniowski:2008qk}
W. Broniowski et~al.,
\newblock Phys. Rev. C80 (2009) 034902.

\bibitem{Bozek:2010aj}
P. Bo\.zek and I. Wyskiel-Piekarska,
\newblock Phys. Rev. C83 (2011) 024910.

\bibitem{Ryblewski:2012rr}
R. Ryblewski and W. Florkowski,
\newblock Phys.Rev. C85 (2012) 064901.

\bibitem{Mauricio:2007vz}
M. Martinez and M. Strickland,
\newblock Phys. Rev. Lett. 100 (2008) 102301.

\bibitem{Schenke:2006yp}
B. Schenke and M. Strickland,
\newblock Phys. Rev. D76 (2007) 025023.

\bibitem{Bhattacharya:2008mv}
L. Bhattacharya and P. Roy,
\newblock Phys. Rev. C79 (2009) 054910.

\bibitem{Strickland:2011aa}
M. Strickland and D. Bazow,
\newblock Nucl.Phys. A879 (2012) 25.

\end{thebibliography}

\end{document}